\documentclass[%
 reprint,
 amsmath,amssymb,
 aps,
]{revtex4-1}

\usepackage{graphicx}
\usepackage{dcolumn}
\usepackage{bm}

\begin{document}

\title{Effects of viscoelasticity on shear-thickening in dilute 
suspensions in a viscoelastic fluid}%

\author{Yuki Matsuoka\(^{ab}\)}
\email{ymatsuoka@sumibe.co.jp}
\author{Yasuya Nakayama\(^{b}\)}
\email{nakayama@chem-eng.kyushu-u.ac.jp}
\author{Toshihisa Kajiwara\(^{b}\)}

\affiliation{%
\(^{a}\)
Corporate Engineering Center, Sumitomo Bakelite Co., Ltd., Shizuoka 426-0041, Japan}%
\affiliation{%
\(^{b}\)Department of Chemical Engineering,
Kyushu University,
Nishi-ku,
Fukuoka 819-0395,
Japan
}%

\date{\today}

\begin{abstract}
We investigate previously unclarified effects of fluid elasticity 
on shear-thickening in dilute suspensions in an Oldroyd-B 
viscoelastic fluid using a novel direct numerical simulation 
based on the smoothed profile method. 
Fluid elasticity is determined by the Weissenberg number $Wi$ and 
by viscosity ratio $1-\beta=\eta_p/(\eta_s+\eta_p)$ which 
measures the coupling between the polymer stress and flow: 
$\eta_p$ and $\eta_{s}$ are the polymer and solvent viscosity, 
respectively.
As $1-\beta$ increases, while the stresslet does not change 
significantly compared to that in the 
$\beta\to 1$ limit, the growth rate of the normalized polymer 
stress with $Wi$ was suppressed.
Analysis of flow and conformation dynamics around a particle for 
different $\beta$ reveals that at large $1-\beta$, polymer stress 
modulates flow, leading to suppression of polymer stretch.
This effect of $\beta$ on polymer stress development indicates 
complex coupling between fluid elasticity and flow, and is 
essential to understand the rheology and hydrodynamic 
interactions in suspensions in viscoelastic media.
\end{abstract}

\maketitle

\section{Introduction}
Particle suspensions in viscoelastic fluids, such as 
polymer solutions or polymer melts, are widely used in industrial 
products. 
To handle such suspensions effectively and efficiently, understanding their rheology is essential.
However, the influence of the media's viscoelasticity on suspension rheology is not well understood. 
One type of viscoelastic fluids, so-called Boger fluids, are widely utilized 
to examine the effect of fluid elasticity due to constant shear viscosity. 
Experimental studies have reported that the apparent viscosity of 
suspensions in Boger fluids shear-thickens even at dilute  
particle volume fractions~\cite{Zarraga2001,Scirocco2005,Tanner2013}, 
suggesting complex interactions between the viscoelastic medium and suspended particles.
Scirocco~\textit{et al.}~\cite{Scirocco2005} reported the thickening at $\phi_p=0.068$.
Zarraga~\textit{et al.}~\cite{Zarraga2001} and Dai~\textit{et al.}~\cite{Tanner2013}
did not explicitly mention the thickening at low $\phi_p$ conditions. However, their data at $\phi_p=0.3$~\cite{Zarraga2001} and $\phi_p=0.05$~\cite{Tanner2013} indicates mild shear-thickening.

Recently, to understand the mechanisms of this 
thickening, the rheology of a dilute suspension in an Oldroyd-B fluid 
have been studied theoretically and 
numerically~\cite{Koch2016,Yang2016,Einarsson2018,Yang2018,Yang2018a}. 
The Oldroyd-B fluid is one of the simplest constitutive models of viscoelastic fluids such as polymer solutions~\cite{Bird1987}. The shear viscosity and the first normal stress difference~(NSD)
coefficient of an Oldroyd-B fluid are independent of the applied 
shear rate $\dot{\gamma}$.
This rate-independent shear viscosity combined with finite elasticity is desirable for modeling the steady shear behavior of Boger fluids.
Although the Oldroyd-B fluid can not capture the whole rheological behavior of real viscoelastic fluids due to it's simplicity~\cite{James2009}, from another point of view, it's simplicity is helpful to obtain a fundamental insight into the separate effects of elasticity and shear viscosity.
Elasticity of Oldroyd-B fluids is characterized by two parameters: 
relaxation time $\lambda$ and viscosity 
ratio $\beta=\eta_s/\eta_0$, where $\eta_s$ and $\eta_p$ are the 
viscosity of the solvent and polymer, respectively, and 
$\eta_0=\eta_s+\eta_p$ is the zero-shear viscosity.
By definition, \(0\leq\beta\leq 1\).
Here, a small \(\beta\) corresponds to a high polymer concentration or 
a high-molecular-weight polymer, indicating strong fluid elasticity~\cite{Bird1987}.
The Weissenberg number $Wi=\lambda\dot{\gamma}$ measures the 
viscoelasticity strength under an applied rate of $\dot{\gamma}$ irrespective of $\beta$.
Koch \textit{et al.}~\cite{Koch2016} and Einarsson \textit{et al.}~\cite{Einarsson2018} applied the perturbation theory and 
demonstrated that the suspension viscosity of Oldroyd-B fluid shear thickens. 
Yang~\textit{et al.} numerically simulated a previously reported system~\cite{Einarsson2018} 
and concluded that the particle-induced fluid stress around the particles is the 
primary source of suspension shear-thickening~\cite{Yang2016,Yang2018}. 

In studies about flow-induced particle clustering in 
viscoelastic fluids, the fluid elasticity is characterized by 
the elastic parameter $S_R=N_1/(2\sigma_{xy})$, where $N_1$ and 
$\sigma_{xy}$ are the first NSD and the shear stress, respectively~\cite{Scirocco2004,Won2004,Pasquino2010,Hwang2011,SantosdeOliveira2011,SantosdeOliveira2012,Choi2012,Pasquino2013,VanLoon2014,Pasquino2014,Jaensson2016}.
For Oldroyd-B fluids, $S_R=(1-\beta)Wi$, which suggests that 
not only $Wi$ but also $1-\beta$
increase the clustering tendency.
The question is whether this trend can also explain the 
shear-thickening of suspensions in viscoelastic fluids; 
do both $1-\beta$ and $Wi$ enhance the shear-thickening?
While the positive effect of $Wi$ on shear-thickening has been 
revealed in recent studies~\cite{Koch2016,Yang2016,Einarsson2018,Yang2018,Yang2018a}, the detailed effect of $\beta$ remains unclear.
The theories by Koch~\textit{et al.}~\cite{Koch2016} and Einarsson~\textit{et al.}~\cite{Einarsson2018} are perturbation theories with the polymer concentration and $Wi$, respectively. Thus at high $S_R$ conditions where $1-\beta$ and $Wi$ are both large, these theories are inadaptable. To evaluate the nonlinear suspension behavior at large $S_R$ conditions,
we need to conduct numerical calculations which fully solve the governing equations.
The numerical studies by Yang and Shaqfeh~\cite{Yang2018} mainly focused on the 
thickening mechanism at $\beta\to1$ condition, 
where the feedback of the polymer stress to the flow can be ignored, i.e. flow field is not perturbed by the polymer stress. In such extreme conditions, since the polymer stress can be analyzed separately from the flow field, analytic perturbation theories have been developed~\cite{Koch2016,Einarsson2018}, and then examined by numerical calculations~\cite{Yang2018}.
However, real viscoelastic fluids used for industrial purposes show the finite polymer concentrations, where the coupling between the polymer stress and flow represented by $\beta$ value should be more important. Therefore, to understand the thickening mechanism in general viscoelastic suspensions, the effects of $\beta$ need to be clarified.
Yang and Shaqfeh showed the change in the $Wi$-dependence of the shear-thickening in polymer stress at a moderate $\beta$ value of 0.68, and only
mentioned the effects of flow modulation by large polymer stress\cite{Yang2018}; however, the underlying $\beta$ dependence of flow and polymer stress was not analyzed. 
For situations where the polymer stress and the flow field strongly couples, physics of shear-thickening in viscoelastic suspensions was not fully explored.
On these backgrounds, the purpose of this article is to clarify 
the effects of fluid's viscoelasticity, specifically the coupling 
between the polymer stress and flow in a wide range of $Wi$, on 
the shear-thickening of the suspensions in Oldroyd-B fluids.
We first briefly explain our newly developed numerical method. 
Then, we present the calculation results of suspension viscosity and NSD coefficient, which indicate the non-trivial effect of $\beta$ on shear-thickening. 
Finally, the mechanism of this effect is investigated through 
interactions between stress and flow fields around the particles.

\section{Simulation method}

\subsection{Governing equations}
Several direct numerical simulations (DNS), which use the fluid 
mesh independent of the surface boundaries of particles rather than body-fitted mesh~\cite{Ahamadi2008,Ahamadi2010,Choi2010,Choi2012,Jaensson2015,Jaensson2016,Yang2016,Yang2018}, 
and particle based methods, which express a viscoelastic fluid as discrete fluid particles, have been adapted for 
suspensions in a viscoelastic fluid in 
2D~\cite{Hwang2004,Hwang2011,Pasquino2014} and 
3D space~\cite{SantosdeOliveira2011,SantosdeOliveira2012,DAvino2013,Vazquez-Quesada2017,Krishnan2017,Yang2018a,Vazquez-Quesada2019}. 
One of the authors proposed the smoothed profile method~(SPM), 
an efficient DNS for suspensions in which interaction 
between particles and the medium is treated through the smoothed 
profile function~\cite{Nakayama2005,Nakayama2008}.  
In SPM, regular mesh rather than body-fitted mesh can be used; therefore, 
the calculation cost of fluid fields, which is dominant in total calculation costs, is nearly independent of the particle number\cite{Nakayama2008}.
This is advantageous for simulation of dense suspensions that containing many particles.
For examples, using SPM, the shear viscosity~\cite{Iwashita2009,Kobayashi2011,Molina2016} and complex modulus~\cite{Iwashita2010} and particle coagulation rate~\cite{Matsuoka2012} 
of Brownian suspensions up to $\phi_p \leq 0.56$ in Newtonian fluids were 
efficiently evaluated. 
Application of SPM was extended to complex host fluids, such 
as electrolyte solutions~\cite{Kim2006,Nakayama2008,luo2010modeling} and to active swimmer suspensions~\cite{Molina2013}. 
Since it can be applied to any continuum solvers, SPM combined with the lattice-Boltzmann method for a 
viscoelastic fluid has been reported~\cite{Lee2017}. 
In this study, we developed a DNS with SPM 
that efficiently evaluates the bulk rheology of suspension in a viscoelastic 
fluid in 3D space.

We consider the suspension of neutrally buoyant and non-Brownian $N$ 
spherical particles with radius $a$, mass $M_p$, and 
moment of inertia $\bm{I}_p$ in a viscoelastic fluid. 
Hereafter unless otherwise stated, all the physical quantities 
are non-dimensionalized by length unit $a$, velocity unit 
$a\dot{\gamma}$, and stress unit $\eta_0\dot{\gamma}$. Hence, 
$M_p$ and $\bm{I}_p$ are non-dimensionalized as 
$\tilde{M}_p=M_p\dot{\gamma}/(\eta_0a)$ and 
$\tilde{\bm{I}}_p=\bm{I}_p\dot{\gamma}/(\eta_0a^3)$, respectively. 

Non-dimensional velocity 
field $\bm{u}(\bm{r}, t)$ at position $\bm{r}$ and time $t$ is 
governed as follows:
\begin{eqnarray}
\label{eq:navier-stokes}
Re\left(\frac{\partial}{\partial t}+\bm{u}\cdot\nabla\right)\bm{u}&=&\nabla\cdot(\bm{\sigma}_n+\bm{\sigma}_p)+Re\phi\bm{f}_p,\\
\label{eq:cnt_eq}
\nabla\cdot\bm{u}&=&0,
\end{eqnarray}
where $Re=\rho a^2\dot{\gamma}/\eta_0$, $\rho$, $\bm{\sigma}_n=-p\bm{I}+2\beta\bm{D}$, $\bm{\sigma}_p$, 
$\bm{I}$, $\bm{D}=(\nabla\bm{u}+\nabla\bm{u}^T)/2$, and $p$ 
are the Reynolds number, fluid mass density, Newtonian solvent stress, and polymer stress, unit tensor, 
strain-rate tensor, and pressure, respectively.
In SPM, The particle profile field is introduced as 
$\phi(\bm{r},t)\equiv\sum_{i=1}^N\phi_i$, where $\phi_i\in[0,1]$ 
is the $i$th particle profile function having continuous diffuse 
interface with thickness $\xi$ and indicating the inside and outside 
of particles by $\phi=0$ and $\phi=1$, respectively~\cite{Nakayama2008}. 
The total velocity field $\bm{u}$ is given by
\begin{equation}
\bm{u}(\bm{r},t)=(1-\phi)\bm{u}_f+\phi\bm{u}_p
\end{equation}
where $\bm{u}_f$ and $\bm{u}_p$ are the fluid and particle velocity fields, respectively.
The body force $\phi\bm{f}_p$ enforces particle rigidity in the velocity field, 
which is defined in the temporal discretization of Eq.~(\ref{eq:navier-stokes})~\cite{Nakayama2008,Molina2016}. The time-integrated body force $\phi\bm{f}_p$ is calculated as
\begin{equation}
\int_t^{t+\Delta t}\phi\bm{f}_pds=\phi(\bm{u}_p-\bm{u}^*)    
\end{equation}
where $\Delta t$ is the simulation time step and $\bm{u}^*$ is the adjacent intermediate total velocity field updated using Eq.~(\ref{eq:navier-stokes}) without the last term in the fractional time stepping.

The individual particles evolve by 
\begin{eqnarray}
\dot{\bm{R}}_i&=&\bm{V}_i,\label{eq:newton}\\
\tilde{M}_p\dot{\bm{V}}_i&=&\bm{F}^H_i+\bm{F}^C_i,\label{eq:newton2}\\
\tilde{\bm{I}}_p\cdot\dot{\bm{\Omega}}_i&=&\bm{N}^H_i,\label{eq:eular}
\end{eqnarray}  
where $\bm{R}_i$, $\bm{V}_i$, and $\bm{\Omega}_i$ are the position, velocity 
and angular velocity of the $i$th particle, respectively.
$\bm{F}_i^H,\bm{N}_i^H$~\cite{Nakayama2008,Molina2016} are the hydrodynamic force and torque, respectively, and $\bm{F}_i^C$ is a 
potential force due to the excluded volume that prevents particles from overlapping. 
The hydrodynamic force $\bm{F}_i^H, \bm{N}_i^H$ are determined by Newton's third law of motion:
\begin{eqnarray}
\int_t^{t+\Delta t}\bm{F}_i^Hds&=&\int Re\phi_i(\bm{u}^*-\bm{u}_p^*)d\bm{r},\\
\int_t^{t+\Delta t}\bm{N}_i^Hds&=&\int\bm{r}_i\times Re\phi_i(\bm{u}^*-\bm{u}_p^*)d\bm{r}
\end{eqnarray}
where $\bm{u}_p^*$ is the intermediate particle velocity field freely advected by the previous particle velocity and $\bm{r}_i=\bm{r}-\bm{R_i}$. The particle velocity field $\bm{u}_p$ is calculated as
\begin{equation}
\label{eq:up}
\phi\bm{u}_p=\sum_{i=1}^{N}\phi_i\left[\bm{V}_i+\bm{\Omega}_i\times\bm{r}_i\right].
\end{equation}
The detail computational algorithm about the fractional time stepping is elaborated on in Refs.~\cite{Nakayama2008,Molina2016}. 

For polymer stress, we use Oldroyd-B fluid: 
\begin{eqnarray}
\label{eq:conformation}
\left(\frac{\partial}{\partial t}+\bm{u}\cdot\nabla\right)\bm{C}&=&(\nabla\bm{u})^T\cdot\bm{C}+\bm{C}\cdot(\nabla\bm{u})-\frac{\bm{C}-\bm{I}}{Wi},
\\
\label{eq:polymer_stress}
\bm{\sigma}_p&=&\frac{(1-\beta)(\bm{C}-\bm{I})}{Wi},
\end{eqnarray}
where $\bm{C}(\bm{r},t)$ is the conformation tensor. 
Oldroyd-B fluid microscopically corresponds to a dilute suspension of dumbbells with a linear elastic spring in an Newtonian solvent~\cite{Bird1987}.  
Conformation tensor $\bm{C}$ is expressed as $C_{ij}=\langle X_iX_j\rangle$, where $\bm{X}$ is the dumbbell's end to end vector normalized by the radius of gyration of polymer and $\langle\cdot\rangle$ is the ensemble average.
The average stretch and orientation of dumbbells 
are $\mathrm{tr}\bm{C}-3$ and the major orientation 
of $\bm{C}$, respectively.
The deformation and orientation of $\bm{C}$ determine the polymer stress.
When $\bm{C}=\bm{I}$, polymer stress is zero in the completely relaxed state. For shear stress component, $\sigma_{p,xy}=(1-\beta)/(2Wi)\langle\bm{X}^2\sin{2\theta}\rangle$, where $\theta$ is the dumbbell's orientation angle from the shear flow direction.
As explained later, this dumbbell representation of $\bm{C}$ is effectively 
interpreted using novelly introduced conformation ellipsoid that is constructed by the eigenvalues and eigenvectors of $\bm{C}$. 
The polymer stress modulates the flow through $\nabla\cdot\bm{\sigma}_p$ in Eq.~(\ref{eq:navier-stokes}).
Eventually, the balance between viscous and polymer stresses, 
and external shear driving results in the steady state.

\subsection{Stress calculation}
The instantaneous volume-averaged stress of the suspension is evaluated in SPM~\cite{Nakayama2008,Iwashita2009,Molina2016} by
\begin{equation} 
\label{eq:spm_eq}
\bm{\sigma}^{\rm sus}=\frac{1}{V}\int_V[\bm{\sigma}_n+\bm{\sigma}_p-\bm{r}Re\phi\bm{f}_p+\bm{ru}\cdot\nabla(Re\bm{u})]d\bm{r},
\end{equation}
where $V$ is the total volume of system and the last term on the right hand side comes from the convective momentum-flux tensor, which is negligible on time averaging over the steady state~\cite{Iwashita2009}.
By assuming ergodicity, the ensemble average of the stress $\langle\bm{\sigma}^{\rm sus}\rangle$ is equated to the average over time.

In suspension rheology, the stress decomposition has been utilized for the evaluation of each contribution of stress components.
In this study, we adopt the procedure proposed by Yang \textit{et al.}~\cite{Yang2016} as follows,
\begin{eqnarray}
\langle\bm{\sigma}^{\rm sus}\rangle&=&\langle\bm{\sigma}^{F0}\rangle+\frac{N}{V}(\langle\bm{\Sigma}\rangle+\langle\bm{S}\rangle),\\
\label{eq:Sigma}
\bm{\Sigma}&=&\frac{1}{N}\int_V(\bm{\sigma}^F-\bm{\sigma}^{F0})\mathrm{d}\bm{r},\\
\label{eq:stresslet}
\bm{S}&=&\frac{1}{N}\int_{S_p}(\bm{r}(\bm{n}\cdot\bm{\sigma}^F))^{sym}\mathrm{d}S,
\end{eqnarray} 
where $\bm{\sigma}^F$ is the stress in the fluid region, 
$\bm{\sigma}^{F0}$ is the fluid stress  without particles 
under the simple shear flow, $S_p$ is the surface of particles, and $(\bm{A})^{sym}$ denotes the symmetric part of tensor $\bm{A}$. 
$\bm{\Sigma}$ represents the stress induced by one particle inclusion
in the fluid region, and $\bm{S}$ is the stresslet.
Evaluation of Eqs.~(\ref{eq:Sigma})-(\ref{eq:stresslet}) requires surface or volume integrals. 
To calculate these integrals numerically in the immersed boundary method,
appropriate location of the particle-fluid interface should be carefully examined~\cite{Yang2018a}. 
By contrast, in SPM, due to the diffuse interface of the smoothed profile function, 
the integrals in Eqs.~(\ref{eq:Sigma})-(\ref{eq:stresslet}) are simply evaluated as follows,
\begin{eqnarray}
\bm{\Sigma}&\approx&\frac{1}{N}\int_V[(\bm{\sigma}_n+\bm{\sigma}_p)^F-(\bm{\sigma}_n+\bm{\sigma}_p)^{F0}]d\bm{r} ,\\
\bm{S}&\approx&-\frac{1}{N}\int_V\bm{r}Re\phi\bm{f}_pd\bm{r}.
\end{eqnarray}
The numerical results by our simple formalism of stress decomposition reasonably agree with those by body-fitted mesh method~\cite{Yang2018} as seen later.

For convenience, viscometric functions are non-dimensionalized as, $\eta\equiv\langle\sigma^{\rm sus}_{xy}\rangle$ and $\Psi_1\equiv\langle\sigma^{\rm sus}_{xx}-\sigma^{\rm sus}_{yy}\rangle/Wi$, and are also decomposed to each contributions as follows~\cite{Yang2016,Yang2018,Yang2018a},
\begin{eqnarray}
\label{eq:eta_decomp}
\eta&=&\eta^0+\frac{N}{V}\eta^p,\\
\label{eq:psi_decomp}
\Psi_1&=&\Psi_1^0+\frac{N}{V}\Psi_1^p,
\end{eqnarray}
where $\eta^0\equiv\langle\sigma_{xy}^{F0}\rangle$ and $\Psi_1^0\equiv\langle\sigma_{xx}^{F0}-\sigma_{yy}^{F0}\rangle/Wi$ are the non-dimensional fluid viscosity and first NSD coefficient without particles.
Note that the NSD coefficient is normalized by $\eta_0\lambda$ rather than $\Psi_{1,0}=2\eta_p\lambda$, which is the NSD coefficient of a Oldroyd-B fluid without particles, 
since our units of the stress and rate are $\eta_0\dot{\gamma}$ and $Wi$, respectively.
$\eta^p$ and $\Psi_1^p$ are the particle contributions to the suspension viscosity and first NSD coefficient, respectively;
\begin{eqnarray}
\label{eq:eta_p_decomp}
\eta^p&\equiv&\langle\Sigma_{xy}\rangle+\langle S_{xy}\rangle,\\
\label{eq:psi_p_decomp}
\Psi_1^p&\equiv&\frac{\langle\Sigma_{xx}-\Sigma_{yy}\rangle+\langle S_{xx}-S_{yy}\rangle}{Wi},
\end{eqnarray}
where $\bm{\Sigma}$ and $\bm{S}$ are non-dimensionalized by $\eta_0\dot{\gamma}a^3$.

\begin{figure}[thb]
\includegraphics[scale=1.1]{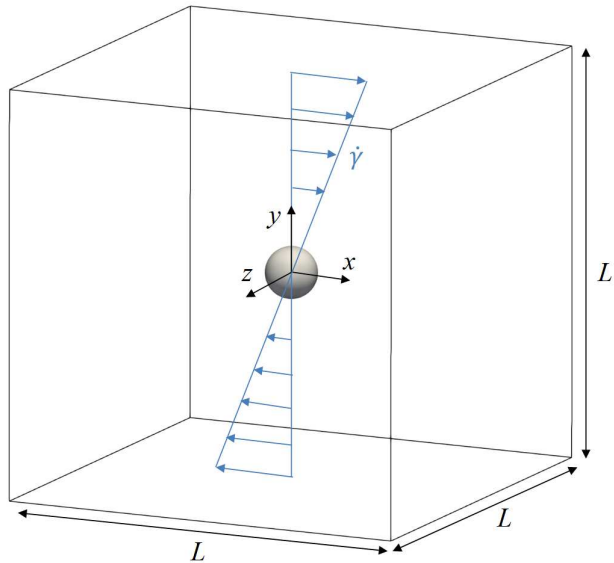}
\caption{\label{system} 
Schematic for the simulated system. The shear flow, shear gradient and vorticity directions are along with Cartesian x,~y,~z-axes, respectively. Periodic boundary conditions are imposed on all faces of the cubic system. 
}
\end{figure}

\subsection{Numerical implementation}

To impose the simple shear flow on the system, we use the 
time-dependent oblique coordinate evolving with mean shear 
velocity as
$\boldsymbol{r}(t)=\boldsymbol{r}(0)-\dot{\gamma}ty\boldsymbol{e}_x$ 
and solve evolution equations (Eqs.~(\ref{eq:navier-stokes})-(\ref{eq:cnt_eq}),~(\ref{eq:conformation})-(\ref{eq:polymer_stress})) formulated on the moving coordinate~\cite{Luo2004,Venturi2009,Kobayashi2011,Molina2016}, where $\bm{e}_x$ is a Cartesian x-axis basis vector. 
The particle equations (Eqs.~(\ref{eq:newton})-(\ref{eq:eular})) are solved under Lees-Edwards boundary conditions~\cite{Kobayashi2011,Molina2016}.
This formulation enables us to impose the full periodic boundary condition and evaluate the bulk rheological properties of the suspension without wall effects. 
The similar full periodic boundary conditions were adopted for steady shear 2D simulations~\cite{Hwang2004,Jaensson2015} and dynamic shear 3D simulations~\cite{DAvino2013} of viscoelastic suspensions.
To the best of our knowledge, this study is the first to report the steady shear simulation of a full periodic 3D system of viscoelastic suspensions.

The evolution equations are solved using the spectral method, which naturally matches the full periodic boundary condition.
The stability condition given by the momentum diffusion term is adopted for determining the simulation time step; $\Delta t=\rho/\eta_0K_{max}^2$ ($K_{max}$ is the largest wave number in our spectral scheme).

\section{Results and discussion}

\subsection{Simulation conditions}
In this study, we focus on a dilute suspension condition. 
As Figure~\ref{system} shows, one particle is located at the center of the cubic box 
$[-L/2,L/2]^3$, where $L=128\Delta$ is the system length and $\Delta$ is the lattice length. 
The mesh resolutions of particles are $a=8\Delta$ and $\xi=2\Delta$. 
In this setup, the particle volume fraction $\phi_p=0.00102$.
This very dilute condition is hardly achieved experimentally. However, such dilute condition, where the complex inter-particle effects are negligible, is preferable to examine the fundamental effect of the interaction between the medium and one particle.
Now that we treat only one particle ($N=1$), the inter-particle force $\bm{F}^C$ in Eq.~(\ref{eq:newton}) can be ignored.
Simple shear flow is imposed on the whole system and then the viscometric functions of suspensions at steady states are evaluated.

\begin{figure}[t]
	\includegraphics[scale=1]{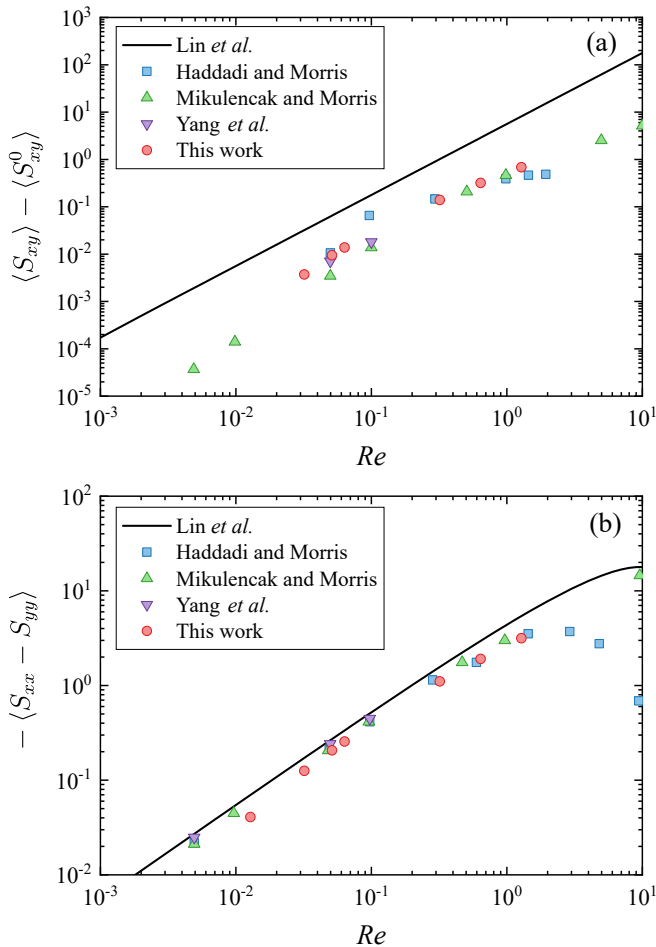}
	\caption{\label{newton_stresslet} 
		Each component of stresslets in a dilute suspension in a Newtonian medium as a function of $Re$: 
		(a) shear component $\langle S_{xy}\rangle$ (b) the first NSD components $-\langle S_{xx}-S_{yy}\rangle$. 
		In Fig.~\ref{newton_stresslet}(a), the results at $Re=0.0128$ is used as the value of $\langle S_{xy}^0\rangle$.
		The past numerical simulation results~\cite{MIKULENCAK2004,Haddadi2014,Yang2016} and theoretical lines~\cite{LIN1970} are also displayed. 
	}
\end{figure}

\subsection{Validation of SPM for sheared viscoelastic suspension}
In order to validate the SPM, we calculate the shear stress and NSD for $\beta=0.99$ case. For evaluation the stress of a viscoelastic suspension, 
inertial contribution to the stress is calculated first.
Although the Reynolds number is set small ($Re\leq0.0633$) to avoid the inertial effect, the inertial effect at a finite Reynolds number can not be ignored especially with the NSD components of suspension stresslets 
because both the inertial and non-inertial contributions to the NSD components are comparable.
To remove finite inertial effects from viscometric evaluations, we follow the procedure proposed by Yang  \textit{et al.}~\cite{Yang2016}.
Figure~\ref{newton_stresslet} shows the Reynolds number dependence of each stresslet component in a Newtonian suspension calculated at the same system shown in Fig.~\ref{system}.
Our results agree well with the past numerical results~\cite{MIKULENCAK2004,Haddadi2014,Yang2016}.
In the evaluation for NSD of viscoelastic suspensions in the followings, the inertial contributions of stresslets at the corresponding Reynolds number conditions in Fig~\ref{newton_stresslet} are subtracted from the results of viscoelastic suspensions.

\begin{figure}[t]
\includegraphics[scale=1]{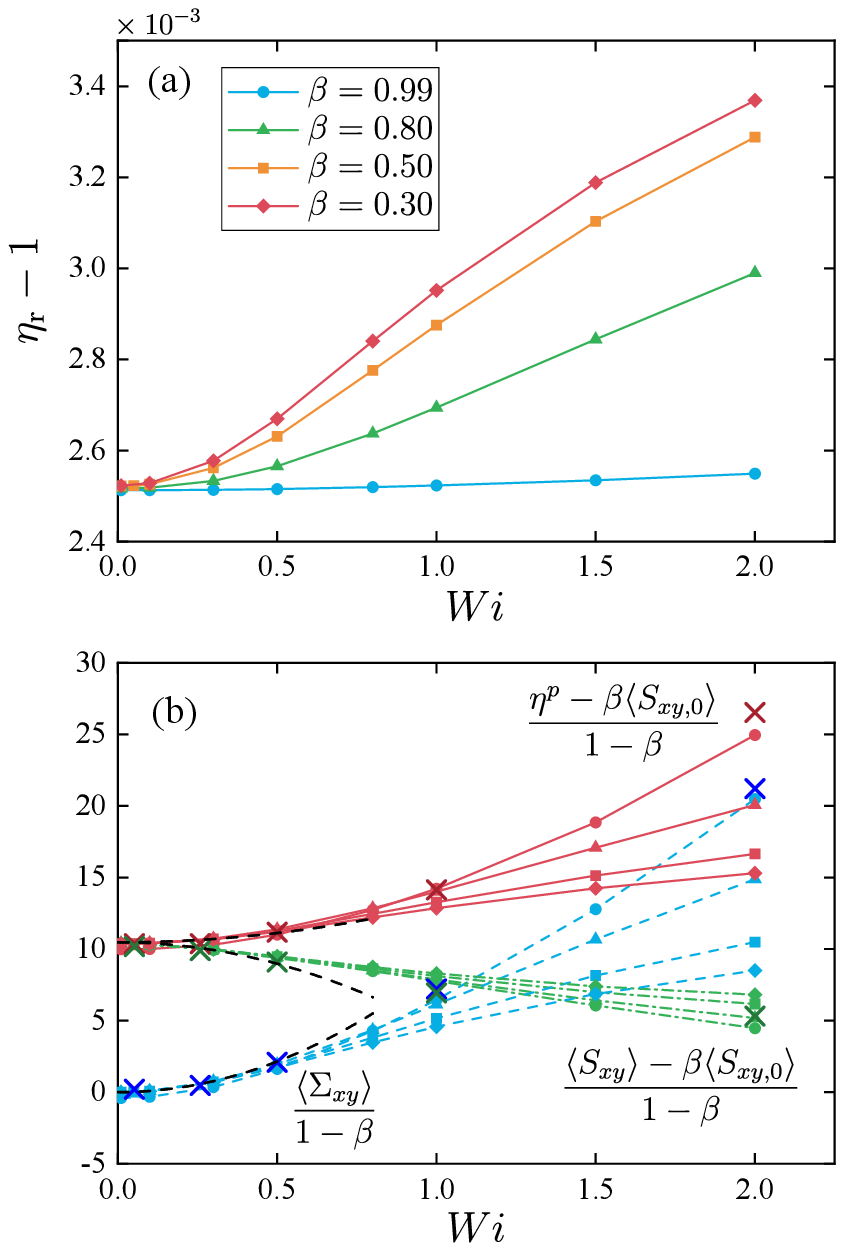}
\caption{\label{fig1} 
Suspension viscosity as a function of 
$\beta$ and $Wi$: (a) normalized viscosity $\eta_r$ (b) 
shear stress decomposition: stresslet (green), fluid stress (blue), 
stresslet+fluid stress (red). 
$\langle S_{xy,0}\rangle$ is the stresslet by Newtonian contribution at $Wi=0$.
Different symbols are for $\beta=0.3(\blacklozenge), 0.5(\blacksquare), 0.8(\blacktriangle), 
0.99(\bullet)$. 
The crosses are from Yang and Shaqfeh's DNS results ($\beta=0.99$)~\cite{Yang2018}, and the dashed-line
is the perturbation solution by Einarsson \textit{et al.} ($\beta=0.99$)~\cite{Einarsson2018}.
}
\end{figure}

Next, the shear stress and first NSD for $\beta=0.99$ is examined. 
Comparison is reported in Fig.~\ref{fig1}(b) for normalized viscosity and 
Fig.~\ref{Psi_1r}(b) for normalized first NSD coefficient. 
Our viscosity for $\beta=0.99$ is in good agreement with theoretical~\cite{Einarsson2018} and numerical~\cite{Yang2018} results (Fig.~\ref{fig1}(b)). Our first NSD coefficient for $\beta=0.99$ also agrees with theoretical~\cite{Koch2006,Greco2007} and numerical~\cite{Yang2018} results (Fig.~\ref{Psi_1r}(b)). 
In the following, we discuss the cases with the strong flow-polymer stress coupling represented by finite $1-\beta$.

\begin{figure}[t]
	\includegraphics[scale=1.1]{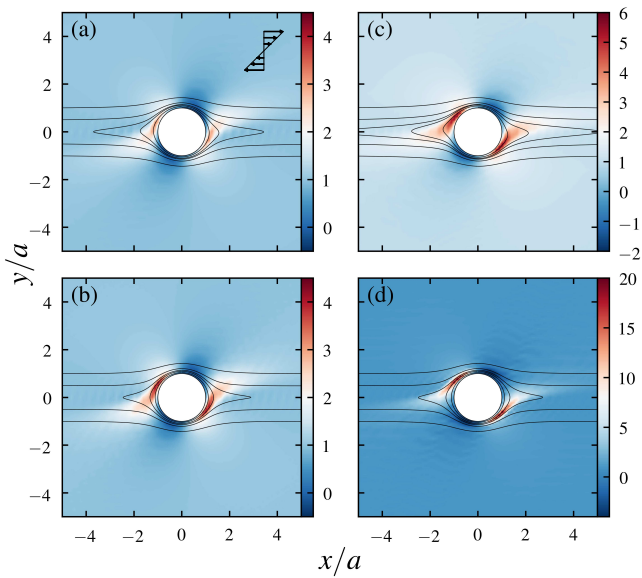}
	\caption{\label{fig2} (Color online) 
		Distribution of shear component of normalized polymer 
		stress $\sigma_{p,xy}/(1-\beta)$ on shear plane~($xy$-plane) 
		through the center of a particle. 
		Positive contribution to shear-thickening is indicated
		by $\sigma_{p,xy}/(1-\beta)>1$ where 
		the polymer shear stress is higher than $\eta_p\dot{\gamma}$. 
		Solid lines show the streamlines. 
		(a)~$\beta=0.5,Wi=1$; (b)~$\beta=0.99,Wi=1$; (c)~$\beta=0.5,Wi=2$; (d)~$\beta=0.99,Wi=2$.}
\end{figure}

\subsection{The effects of $\beta$ on the suspension viscosity}
Figure~\ref{fig1}(a) shows the relative shear viscosity of a suspension 
$\eta_r=\eta=\langle\sigma_{xy}^{\rm sus}\rangle$ to $\eta_0$, as a 
function of $\beta$ and $Wi$.
The relative apparent viscosity $\eta_r$ increases with $Wi$, 
which indicates shear-thickening. 
The shear-thickening is more pronounced for larger $1-\beta$, 
which is attributed to the relative increase of polymer 
stress contribution. 
To further investigate the origin of this shear-thickening, the particle contributions to suspension shear stress 
are decomposed according to Eq.~(\ref{eq:eta_p_decomp}) in Fig.~\ref{fig1}(b).
In Fig.~\ref{fig1}(b), each stress component is additionally normalized by 
$1-\beta$, which represents the ratio of stress to $\eta_p\dot{\gamma}$ 
since now the normalization unit is $(1-\beta)\eta_0\dot{\gamma}=\eta_p\dot{\gamma}$.
As $Wi$ increases, while $\langle S_{xy}\rangle$ shear-thins, $\langle\Sigma_{xy}\rangle$ 
strongly shear thickens; thus, summing them yields the total shear-thickening. 
Relative to the $\beta$ dependence shown in Fig.~\ref{fig1}(b), 
we observe a non-trivial trend, i.e., thickening rate of $\langle\Sigma_{xy}\rangle/(1-\beta)$ 
for $Wi \gtrsim 1$ weakens at smaller $\beta$. 
This trend can also be observed in Fig.~\ref{fig1}(a) as slower
growth of \(\eta_r\) with $Wi$ at smaller $\beta$. 
These results demonstrate that the growth of fluid elasticity with 
$1-\beta$ weakens the $Wi$-dependence in shear-thickening, 
which indicates that $1-\beta$ and $Wi$ in the elastic parameter 
$S_R$ have counteracting effects on shear-thickening in the 
suspension.

To evaluate the cause of the $\beta$-dependent thickening 
in $\langle\Sigma_{xy}\rangle$, Fig.~\ref{fig2} shows the distribution of 
polymer shear stress around a particle on the shear plane at 
$\beta=0.5$ and $0.99$.
The case $\beta=0.99$ was already analyzed by Yang and Shaqfeh~\cite{Yang2018}, who identified that the main source of $\langle\Sigma_{xy}\rangle$ thickening comes from polymer stress near the particle when the polymer stress is passive to the flow. 
Here, we discuss the effect of $1-\beta$ on the suspension rheology by comparing the cases of $\beta=0.5$ and $0.99$ cases.
We observe localized large shear stress in the recirculation 
region near the particle~(represented by the closed streamlines).
For $Wi <1$, the distributions of the streamline and stress
do not change significantly with $\beta$~(not shown). 
In contrast, for $Wi > 1$, the level of polymer stress concentration
in the recirculation region increases with larger $\beta$.
In other words, the polymer shear stress concentration is 
suppressed with increased fluid elasticity with $1-\beta$.
This change in polymer stress is reflected in the weakening 
of the $Wi$-dependence of shear-thickening in Fig.~\ref{fig1} 
and can be explained by the change in flow caused 
by the polymer stress.
At $\beta=0.5$, the streamline visibly changes with $Wi$ 
(Figs.~\ref{fig2}(a) and \ref{fig2}(c)), which demonstrates strong coupling 
between the polymer stress and the flow, while the 
streamline remains nearly unchanged between $Wi=1$ and $2$ at 
$\beta=0.99$~(Figs.~\ref{fig2}(b) and \ref{fig2}(d)). 
These results clearly indicate that high fluid elasticity at the 
small $\beta$ condition modulates the flow, which suppresses 
local concentration of the polymer stress.

\begin{figure}[t]
\includegraphics[scale=1]{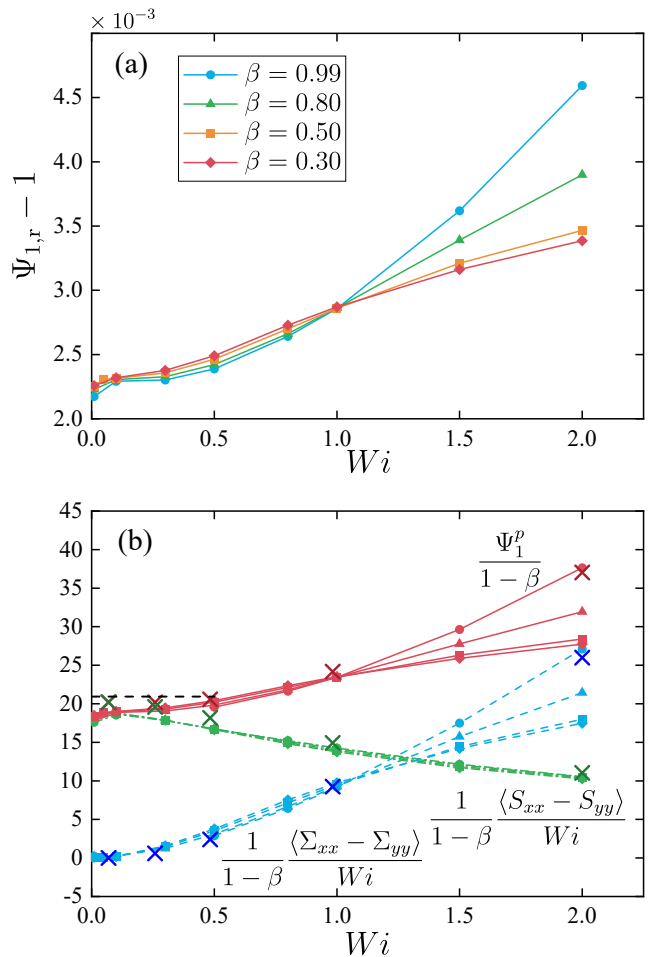}
\caption{\label{Psi_1r} 
Suspension first NSD coefficient as a function of 
$\beta$ and $Wi$: (a) $\Psi_{1,r}$ (b) normal stress decomposition: stresslet (green), fluid stress (blue), stresslet+fluid stress (red). 
Different symbols are for $\beta=0.3(\blacklozenge), 0.5(\blacksquare), 0.8(\blacktriangle), 0.99(\bullet)$. 
The crosses reference Yang and Shaqfeh's DNS results ($\beta=0.99$)~\cite{Yang2018}, and the dashed-line
is the theoretical value of a second order fluid $\Psi_1^p/(1-\beta)=\langle\Sigma_{xx}-\Sigma_{yy}\rangle/((1-\beta)Wi)=20\pi/3$~\cite{Koch2006,Greco2007}.
}
\end{figure}

The shear-thickening and it's changes observed in this study might appear to be minor effects because the viscosity increment of this thickening is small~(Fig.~\ref{fig1}(a)). This is because that the particle concentration in this study is very dilute ($\phi_p=0.001$). 
We note that the shear-thickening in a viscoelastic suspension is qualitatively the result of the coupling between the flow and the viscoelastic response of a medium; modulated flow pattern from simple shear due to particle geometry induces extra viscoelasitic stress that further change the flow around the particle.
This mechanism is supposed to be common in a wide class of viscoelastic suspensions; therefore, this effect is expected to work other systems regardless of the constitutive equation of a medium and/or the type of a particle. In addition, since the viscoelastic stress responsible for the shear-thickening occurs in the vicinity of particles (Fig.~\ref{fig2}), this effect is supposed to be relevant and enhanced with particle concentration even in non-dilute suspensions where inter-particle interaction works.
Furthermore, as explained later, the contribution of elongational flow around a particle (Fig.~\ref{fig3}) suggests that the shear-thickening can become more prominent in viscoelastic media with strong elongational response.

\subsection{The effects of $\beta$ on the first NSD coefficient}
Figure~\ref{Psi_1r}(a) shows the relative first NSD coefficient of a suspension $\Psi_{1,r}=\Psi_1/(2(1-\beta))=\langle\sigma_{xx}^{\rm sus}-\sigma_{yy}^{\rm sus}\rangle/(2(1-\beta)Wi)$ 
to $\Psi_{1,0}=2\eta_p\lambda$, as a function of $\beta$ and $Wi$. 
As seen in $\eta_r$, $\Psi_{1,r}$ increases with $Wi$. 
The increasing rate of $\Psi_{1,r}$ for $Wi > 1$ weakens at smaller $\beta$.
In contrast to $\eta_r$, the enhancement of the first NSD coefficient for larger $1-\beta$ is not observed.
This is simply because $\Psi_{1,0}$, which is the denominator of $\Psi_{1,r}$, purely originates from polymer stress 
and is also pronounced by the same order, $\Psi_{1,0}\propto\eta_p=\eta_0(1-\beta)$, as $\Psi_1$. 
As was done for the suspension viscosity,  the suspension NSD is decomposed according to Eq.~(\ref{eq:psi_p_decomp}) in Fig.~\ref{Psi_1r}(b).
The overall trend is similar to that in the suspension viscosity; as $Wi$ increases, while $\langle S_{xx}-S_{yy}\rangle$ decreases, $\langle\Sigma_{xx}-\Sigma_{yy}\rangle$ strongly increases.

\begin{figure}[t]
\includegraphics[scale=1.08]{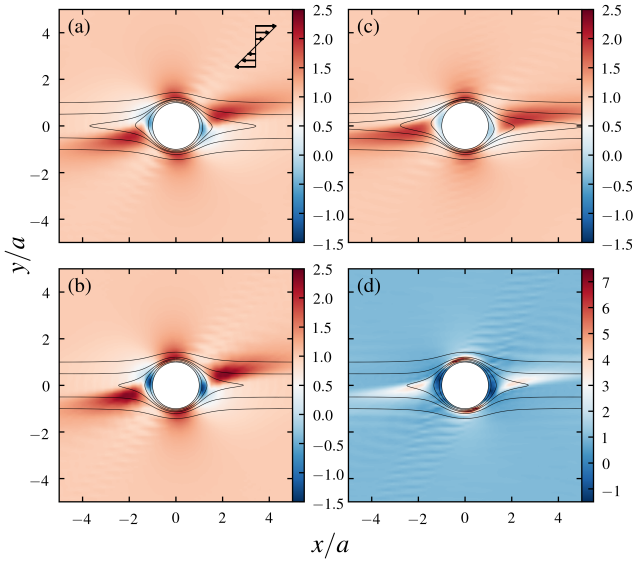}
\caption{\label{N1} (Color online) 
Distribution of NSD component of normalized polymer 
stress $(\sigma_{p,xx}-\sigma_{p,yy})/(2(1-\beta)Wi)$ on shear plane~($xy$-plane) 
through the center of a particle. 
Positive contribution to shear increase is indicated
by $(\sigma_{p,xx}-\sigma_{p,yy})/(2(1-\beta)Wi)>1$ where the polymer NSD is higher than $\Psi_{1,0}\dot{\gamma}^2$. 
Solid lines show the streamlines. 
(a)~$\beta=0.5,Wi=1$; (b)~$\beta=0.99,Wi=1$; (c)~$\beta=0.5,Wi=2$; (d)~$\beta=0.99,Wi=2$.}
\end{figure}

\begin{figure}[t]
\begin{center}
\includegraphics[scale=1.0]{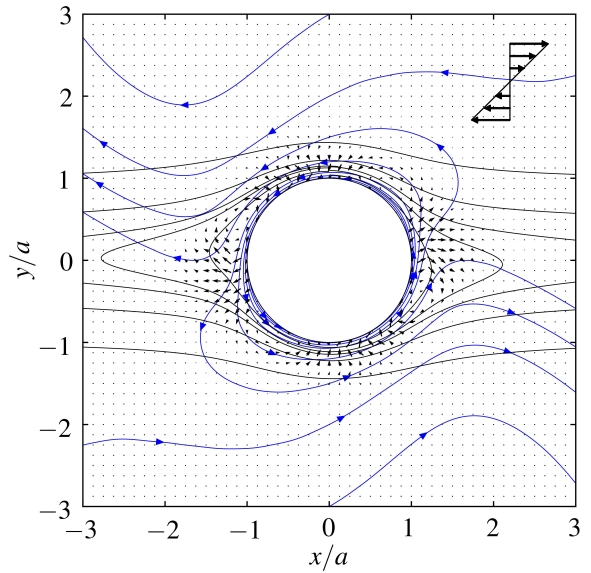}
\caption{\label{disturb} (Color online) 
Modulation of the velocity field at $\beta=0.5$ and $Wi=2$. 
The streamlines are calculated from velocity (black lines) and disturbance velocity $\delta\bm{u}$ (blue lines). The vector field indicates $\nabla\cdot\bm{\sigma}_p$.} \end{center}
\end{figure}

Figure~\ref{N1} shows the distribution of polymer NSD around a particle on the shear plane at $\beta=0.5$ and $0.99$.
The general trend in Fig.~\ref{N1} is the same as that in Fig.~\ref{fig2}.
For $Wi < 1$, the distributions of NSD do not change significantly with $\beta$ (Figs.~\ref{N1}(a) and \ref{N1}(b)).
In contrast, for $Wi > 1$, the level of polymer NSD concentration near the particle surface (especially at the poles of a particle) increases with $\beta$ (Fig.~\ref{N1}(d)).
Note that the high NSD region (Fig.~\ref{N1}) extends more widely to the shear flow direction than the shear stress (Fig.~\ref{fig2}).
For smaller $\beta$ and larger $Wi$ conditions (Fig.~\ref{N1}(c)), high stress region extending outside of recirculation region near the particle surface, where polymers pass through the particle and more align to the flow direction, 
seems to give relatively higher contributions to total polymer NSD increase. 

\begin{figure*}[t]
\begin{center}
\includegraphics[scale=1.15]{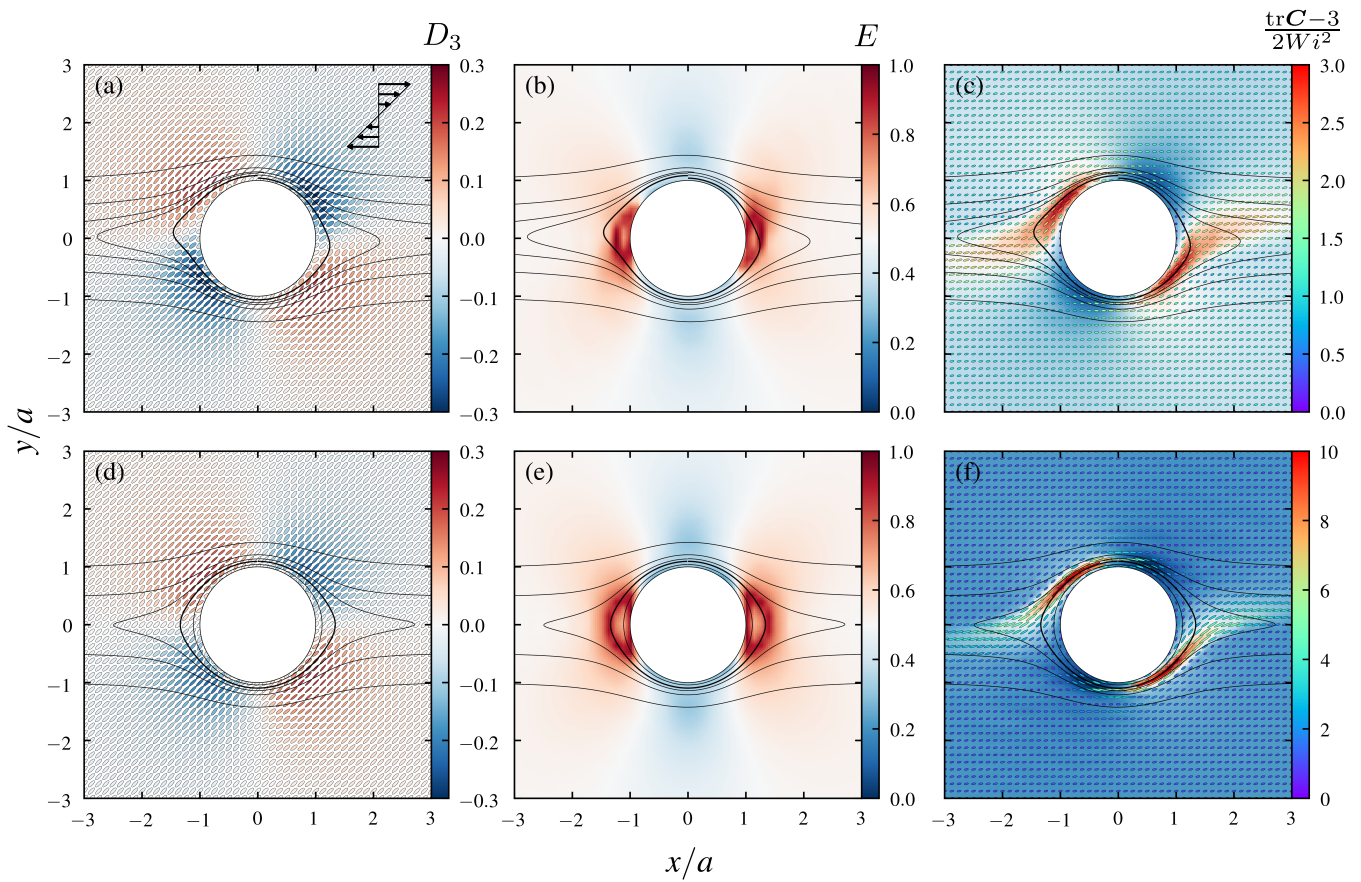}
\caption{\label{fig3} (Color online) 
Distributions of physical quantities at $\beta=0.5$, $0.99$ and $Wi=2$.
The top row panels (a)-(c) and bottom row panels (d)-(f) indicate the results at $\beta=0.5$ and $0.99$, respectively.
(a)(d) ellipsoid from strain rate tensor, 
(b)(e) irrotationality $E$,
(c)(f) ellipsoid from conformation tensor.
In the panels (a)(d), the elipsoid color is from the out-of-plane eigenvalue $D_3$. 
In the panels (c)(f), the background contour indicates the  
normalized polymer shear stress as in Fig.~\ref{fig2}(c)(d), 
and the ellipsoid color indicates normalized polymer stretch $(\mathrm{tr}\bm{C}-3)/(2Wi^2)$, which coincides 
with unity in an Oldroyd-B fluid without particles under simple shear flow.
The bold streamline shows the line that pass through the maximum polymer shear stress.}
\end{center}
\end{figure*}

\subsection{Flow and conformation around a particle}

\begin{figure}[t]
\includegraphics[scale=1]{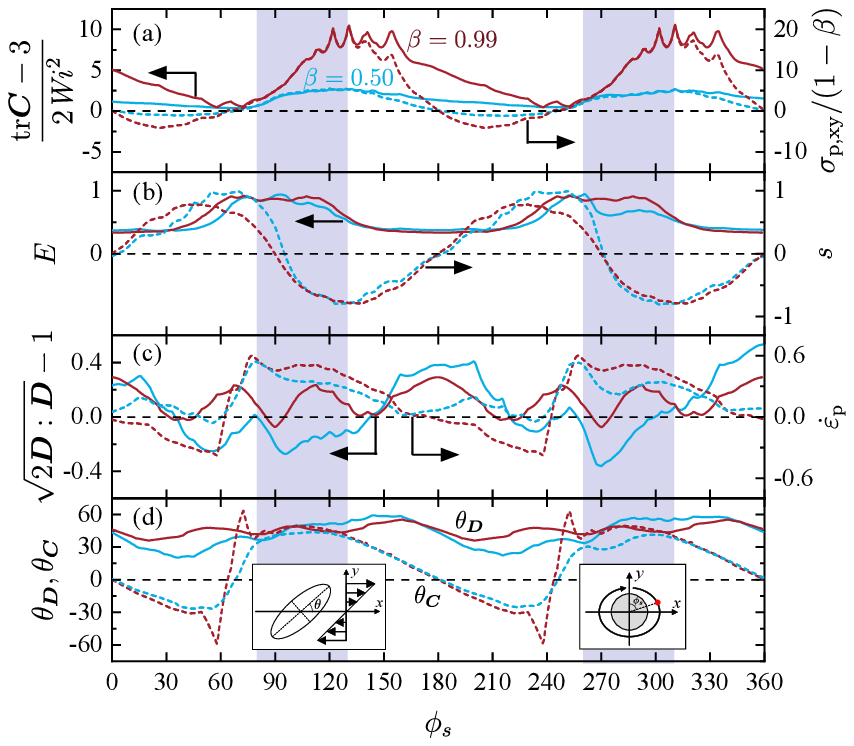}
\caption{\label{fig4} 
Spatial variations of several fields along the streamlines that 
pass through the maximum polymer shear stress at $Wi=2$ shown 
with the bold line in Fig.~\ref{fig3}.
The horizontal axis $\phi_s$ specifies the angle from the 
velocity gradient direction ($y$ axis), depicted in the right 
inset in the panel (d).
The red lines represent $\beta=0.99$; blue lines represent $\beta=0.5$.
With the exception of (d), the solid lines are for left axis, and dashed lines are for right axis.
(a) left: 
$(\mathrm{tr}\bm{C}-3)/(2Wi^2)$, right: 
$\sigma_{p,xy}/(1-\beta)$. 
Spikes at $\beta=0.99$ are artifacts due to mesh resolution. 
(b) left: $E$, right: 
$s=3\sqrt{6}\det\bm{D}/(\bm{D}:\bm{D})^{(3/2)}\in [-1,1]$, which indicates 
the 3D flow patterns~\cite{Nakayama2016}: $s<0$ for biaxial 
elongation, $s=0$ for planar shear, and $s>0$ for uniaxial 
elongation.
(c) left: $\sqrt{2\bm{D}:\bm{D}}-1$, right: $\dot{\varepsilon}_p$ 
(d) $\theta_{\bm{D}},\theta_{\bm{C}}$, the angles of
$\bm{D},\bm{C}$ ellipsoids from the shear direction (left inset in (d)).}
\end{figure}

Here, we focus on the relationship between polymer stress 
and velocity modulation.
Figure~\ref{disturb} shows two streamlines on the shear plane 
at $\beta=0.5$ and $Wi=2$ from the velocity (black line) and 
the disturbance velocity (blue line), i.e., 
$\delta\bm{u}=\bm{u}-\bm{u}_{nw}$, where $\bm{u}_{nw}$ is the 
velocity field in a Newtonian medium at the same $\dot{\gamma}$ 
and $Re$. 
Polymer stress induces the anticlockwise disturbance to the 
velocity near the particle, which corresponds to slowdown of 
particle rotation speed in viscoelastic 
media~\cite{DAvino2008,Snijkers2009,Snijkers2011,Housiadas2011,Housiadas2011b}. 
Furthermore, in the recirculation region, the fluid goes away by 
spiraling out from the particle vicinity to the downstream, thereby 
forming fore-aft asymmetric streamlines, and such phenomena at 
finite $Wi$ have been reported in Second-order Fluid~\cite{Subramanian2007} and Giesekus fluid~\cite{DAvino2008,Housiadas2011} systems.  
Figure~\ref{disturb} also shows the force density vector 
$\nabla\cdot\bm{\sigma}_p$, exhibiting a correlation between  
$\nabla\cdot\bm{\sigma}_p$ and the disturbance streamlines. 
In short, the flow disturbance caused by the polymer stress is consistent with the previously reported flow characteristics in viscoelastic suspensions.

Next, flow pattern and conformation around a particle are analyzed.
Figure~\ref{fig3} indicates the distributions of physical quantities about flow and polymer conformation fields at $\beta=0.5$ (the top row) and $\beta=0.99$ (the bottom row) and $Wi=2.0$.
Figures.~\ref{fig3}(a)(d) and \ref{fig3}(c)(f) show the distributions 
of $\bm{D}$ and $\bm{C}$ with eigen-ellipsoids, respectively.
The eigen-ellipsoid for a symmetric tensor is constructed from 
the three eigenvalues and normalized eigenvectors $(A_i, 
\bm{n}_{i})~(i=1,2,3)$, where we take $A_1>A_2$ and $\bm{n}_3$ is 
directed normal to the shear plane. 
In Figs.~\ref{fig3}(a)(d) and \ref{fig3}(c)(f), the ellipsoids are drawn as their 
major/minor axes are $(1+kA_1)\bm{n}_1$ and $(1+kA_2)\bm{n}_2$, respectively, where $k>0$ is a scaling constant. 

Change in flow pattern around a particle is analyzed with $\bm{D}$-ellipsoid in Figs.~\ref{fig3}(a)(d) and the irrotationality shown in Figs.~\ref{fig3}(b)(e).
The $\bm{D}$-ellipsoid with $D_3$ indicates the flow patterns at 
each position, e.g. an elongated ellipsoid with a negative/positive $D_3$ indicates uniaxial/biaxial elongational flows.
From the upstream to downstream, the flow 
pattern around the particle varies from biaxial elongation, to planar shear, and then uniaxial elongation. 
This flow patterns mainly originate from the existence of the particle: at the upstream, flow avoiding the particle creates biaxial elongational flow while converging flow at the downstream creates uniaxial elongational flow.
Figures~\ref{fig3}(b)(e) display the irrotationality
\begin{equation} 
\label{eq:irrot}
E=\frac{\sqrt{\bm{D}:\bm{D}}}{\sqrt{\bm{D}:\bm{D}}+\sqrt{\bm{\Omega}:\bm{\Omega}^T}},
\end{equation}
 where 
$\bm{\Omega}=(\nabla\bm{u}-\nabla\bm{u}^T)/2$ is the vorticity 
tensor~\cite{Nakayama2016}. 
For rigid-body rotation, $E=0$, while $0<E<1$ for the partially 
rotational flow. Note that the irrotational flow indicated by $E=1$ is 
a strain-dominated flow; thus, $E$ is conveniently used to identify 
the strain-dominated flow because $E$ is normalized.
The appearance of contour of $E$ is similar to that of the 
velocity-gradient eigenvalue discriminant~\cite{Einarsson2018} 
and to that of the second invariant of the velocity gradient~\cite{Yang2018}, 
but the latter quantities are not normalized.
In a very recent work by V\'{a}zquez-Quesada~\textit{et al.}~\cite{Vazquez-Quesada2019}, the dimensionless parameter $Q=(\bm{D}:\bm{D}-\bm{\Omega}:\bm{\Omega}^T)/(\bm{D}:\bm{D}+\bm{\Omega}:\bm{\Omega}^T)$, which is another definition for irrotationality, shows the similar distributions with those of $E$ in our results. Compared with $E$ in Eq.~(\ref{eq:irrot}), $Q$ is defined with a squared Frobenius norm and is normalized to $-1\leq Q\leq 1$. Since both $E$ and $Q$ are functions of $\sqrt{\bm{\Omega}:\bm{\Omega}^T/\bm{D}:\bm{D}}$, they essentially measure the relative magnitude of $\bm{\Omega}$ to that of $\bm{D}$.

The strain-dominated flow develops near the equator of the particle. 
This is because the rotational flow in bulk simple shear flow is hindered in front and backside of the particle.
In contrast, the half-rotational flow develops near the poles 
(Figs.~\ref{fig3}(b)(e)). 
This is because, near the poles, the simple shear flow is recovered.
At $\beta=0.99$, $\bm{D}$-ellipsoids and $E$ around the particle are symmetrically distributed (Figs.~\ref{fig3}(d)(e)). By contrast, at $\beta=0.5$, these distributions are distorted corresponding to the flow modulation (Figs.~\ref{fig3}(a)(b)). 

Change in polymer conformation around a particle is analyzed with $\bm{C}$-ellipsoid in Figs.~\ref{fig3}(c)(f). The shape and orientation of a $\bm{C}$-ellipsoid correspond to the ensemble-averaged stretch and orientation of dumbbell molecules at each position.
In the region with high polymer shear stress, we observe that the 
$\bm{C}$-ellipsoid is highly deformed and oriented by $45^\circ$ 
to the shear flow direction. 
At $\beta=0.99$, the ellipsoids at such region are more stretched than those at $\beta=0.5$ (Fig.~\ref{fig3}(f)). This local high stretching of $\bm{C}$-ellipsoids corresponds to the polymer stress concentration observed in Fig.~\ref{fig2}(d).

The development of polymer stress near the particle is 
determined by the variation of $\bm{C}$-ellipsoids, which is distorted and 
rotated along the recirculation streamline.
Figure~\ref{fig4} shows the variations of fields along 
the recirculation streamline that pass through the maximum polymer shear stress 
at $Wi=2$ (bold streamlines in 
Fig.~\ref{fig3}). 
Along the recirculation, the conformation is periodically 
stretched and relaxed (left axis in Fig.~\ref{fig4}(a)). 
The normalized polymer stretch is proportional to the polymer 
dissipation as
\begin{equation}
\Phi_p=\frac{\mathrm{tr}\bm{\sigma}_p}{2Wi}=(1-\beta)\frac{\mathrm{tr}\bm{C}-3}{2Wi^2}.
\end{equation}
The shear component of the polymer stress (right axis in Fig.~\ref{fig4}(a)) depends on both the polymer stretch and 
the orientation of $\bm{C}$ as denoted by $\theta_{\bm{C}}$ 
(Fig~\ref{fig4}(d)). 
In Fig~\ref{fig4}, the upstream regions of the maxima in 
$\mathrm{tr}\bm{C}$ 
$(\phi_s\in[80^\circ,130^\circ], [260^\circ, 310^\circ])$ 
are colored, at which polymer stretching progresses.
Prior to entering these regions, $E$ (left axis in Fig.~\ref{fig4}(b)) grows immediately, indicating that the flow 
becomes more strain-dominated.
Note that $\theta_{\bm{C}}$ does not change significantly where $E$ is high.
Correspondingly, $\theta_{\bm{C}}$ abruptly approaches $\theta_{\bm{D}}$, the primary direction of $\bm{D}$ 
(Fig.~\ref{fig4}(d)), under uniaxial elongational flow~(right axis in Fig.~\ref{fig4}(b)). 
At $\beta=0.99$, the primary directions of $\bm{D}$ 
and $\bm{C}$ are nearly aligned, i.e., 
$\theta_{\bm{C}}\approx\theta_{\bm{D}}$, in the shaded regions. 
As a result, the flow pattern in the shaded regions in 
Fig.~\ref{fig4} strongly facilitates the polymer stretch with a 
certain level of strain rate $\sqrt{2\bm{D}:\bm{D}}$~(left axis in Fig.~\ref{fig4}(c)) 
under biaxial elongational flow~(right axis in Fig.~\ref{fig4}(b)).
Approaching the poles, $E$ decreases; thus the polymers are 
subject to rotation, which causes a gradual change in 
$\theta_{\bm{C}}$ and a discrepancy between $\theta_{\bm{C}}$ and 
$\theta_{\bm{D}}$.
Due to this misalignment, $\mathrm{tr}\bm{C}$ and the polymer 
shear stress relax regardless of the finite strain rate.

These two effects of strain rate and orientation 
alignment on the polymer stretch are combinedly reflected in 
the effective molecular extension rate
\begin{equation} 
\dot{\varepsilon}_p=\bm{n}_1\bm{n}_1:\bm{D},
\end{equation}
which is the elongation rate in the primary direction of conformation 
$\bm{n}_1$~\cite{Pasquali2002}. 
When the alignment is high, $\dot{\varepsilon}_p$ takes a 
positive value (right axis in Fig.~\ref{fig4}(c)), thereby indicating the 
stretch of the polymers. In contrast, when the orientations are 
rather perpendicular, $\dot{\varepsilon}_p$ takes a negative 
value, which indicates the compression of the polymers.
In the shaded regions in Fig.~\ref{fig4}, polymers are exposed to strong stretching. At $\beta=0.99$, $\dot{\epsilon}_p\approx0.5$. In this situation, the local extension rate normalized by $\lambda$, $Wi\dot{\epsilon}_p=1.0$, is twice larger than 
the strain-hardening threshold of the Oldroyd-B fluid $Wi\dot{\epsilon}_p=0.5$, above which rate, a dumbbell molecule in the Oldroyd-B fluid is subject to unbounded extension~\cite{Bird1987}. This extensional characteristics of the Oldroyd-B fluid is supposed to facilitate the polymer stretch in the shaded region in Fig.~\ref{fig4}(a) and the development of localized large polymer shear stress in Fig.~\ref{fig3}(f).
On the other hand, at $\beta=0.5$, $Wi\dot{\varepsilon}_p$ in the shaded region is reduced to about $0.6$, which reflects the reduced alignment and strain rate for $\beta=0.5$.
This reduction of the local effective extension rate at $\beta=0.5$ results in the weak growth of polymer stretch~(Fig.~\ref{fig4}(a)).
This $\beta$-dependence originates from the modulation of the 
flow at small $\beta$, which causes the changes in the 
flow pattern along the recirculation streamlines and in the 
conformation kinetics.

This analysis explains the polymer stress development near the particle at $\beta=0.99$ and suppression of the stress development at $\beta=0.50$: i.e., the underlying mechanism of the change in the shear-thickening of polymer stress by $\beta$ values, which has not been addressed in the previous studies.
Furthermore, our results suggest that at larger $\beta$ conditions, the $Wi$-dependence of shear-thickening of viscoelastic suspensions is more sensitively affected by the elongational property of the viscoelastic media as pointed out by Yang and Shaqfeh~\cite{Yang2018}.
Conversely, at small $\beta$ conditions, the flow modulation by large polymer stress decreases the local effective extension rate and consequently weakens the $Wi$-dependence of shear-thickening of viscoelastic suspensions.

\section{Conclusions}
In this study, we examined unclear effects of 
fluid elasticity on shear-thickening in the suspension in an 
Oldroyd-B medium with a newly developed direct numerical 
simulation based on the SPM~\cite{Nakayama2005,Nakayama2008,Molina2016}. 
As indicated in the elastic parameter $S_R=(1-\beta)Wi$, 
fluid elasticity is enhanced with both $Wi$ and $1-\beta$.
Our results demonstrate that coupling between the polymer stress and flow 
is enhanced with increasing $1-\beta$, which results in modulation of 
the velocity and the suppression of the increase in the 
normalized polymer stress proximity of the particle.
This change in polymer stress development leads to 
the weakening of the $Wi$-dependence of shear-thickening in 
average polymer stress while the stresslet does not change significantly, 
resulting in non-trivial weakening of the $Wi$-dependence of the total suspension viscometric functions.
Our results suggest that this counteracting effect of fluid 
elasticity with $1-\beta$ and $Wi$ is critically important for 
suspension in real moderate or strongly viscoelastic fluids and hydrodynamic interactions.
Indeed, the weakening of shear-thickening in large $Wi$ region is observed experimentally~\cite{Zarraga2001,Scirocco2005,Tanner2013}, which is consistent with our results.

These results represent a step forward for understanding the role of coupling between viscoelastic response and flow in shear-thickening in viscoelastic suspensions.
Previous experimental studies indicate the weakening of shear-thickening in large $Wi$ region~\cite{Zarraga2001,Scirocco2005,Tanner2013}, 
followed by a numerical study to observe the change in $Wi$ dependence of 
the shear-thickening in average polymer stress at $\beta=0.68$~\cite{Yang2018};
however, how the viscoelastic stress-flow coupling at a finite $1-\beta$ 
alter the shear-thickening was not explored. 
In this study, analysis of flow pattern and conformation kinetics around the particle at a finite $1-\beta$ in addition to $\beta=0.99$ clarified the underlying physics of the change in the bulk suspension rheology by the coupling between the polymer stress and flow, which has not been addressed in the previous studies.
Note that shear-thickening of a suspension can 
also be affected by other factors in constitutive modeling of the 
medium, such as shear thinning, and the extensibility of 
polymers~\cite{Yang2018}. 
Unconstrained extensibility of polymer in Oldroyd-B fluid is most pronounced 
when the polymer stress is passive to flow at $\beta\to1$.
Oldroyd-B fluid used in this study is too simple to capture all the aspects of real viscoelastic fluids and thus our results should be carefully interpreted and validated compared with experimental results in the future.
Nonetheless, the coupling between the polymer stress and the flow observed in this study should be generally relevant regardless of the individual characteristics of different constitutive models.
The detail analysis of flow and conformation in this study would give general insights into microscopic behavior of suspending polymers in real viscoelastic fluids.

\bigskip

\section*{Acknowledgments}
The numerical calculations were mainly carried out using the 
computer facilities at the Research Institute for Information 
Technology at Kyushu University. This work was supported by 
Grants-in-Aid for Scientific Research (JSPS KAKENHI) under Grants 
No.~JP18K03563. Financial support from Hosokawa Powder Technology 
Foundation is also gratefully acknowledged.

\bigskip


%
\end{document}